\documentclass[conference]{IEEEtran}
\ifCLASSINFOpdf
\else
\fi
\usepackage{booktabs}
\usepackage{amsmath,amssymb,amsfonts}
\usepackage{algorithmic}
\usepackage{graphicx}
\usepackage{textcomp}
\usepackage{xcolor}
\usepackage{hyperref}
\usepackage{array}
\usepackage{siunitx}
\usepackage{booktabs}
\usepackage{float}
\usepackage{hyperref}
\usepackage{listings} 
\usepackage{color} 
\usepackage{comment}
\usepackage{multirow}

\usepackage{amsmath}
\usepackage{graphicx}
\usepackage{amsmath}
\usepackage{amsfonts}

\definecolor{codegreen}{rgb}{0,0.6,0}
\definecolor{codegray}{rgb}{0.5,0.5,0.5}
\definecolor{codepurple}{rgb}{0.58,0,0.82}
\definecolor{backcolour}{rgb}{0.95,0.95,0.92}

\usepackage{listings}
\usepackage{xcolor}
\usepackage{array}

\lstdefinestyle{mystyle2}{
    backgroundcolor=\color{backcolour},   
    commentstyle=\color{green},
    keywordstyle=\color{blue},
    stringstyle=\color{purple},
    basicstyle=\ttfamily\footnotesize, 
    breakatwhitespace=false,         
    breaklines=true,                 
    captionpos=b,                    
    keepspaces=true,                 
    numbers=none,                                  
    showspaces=false,                
    showstringspaces=false,
    showtabs=false,                  
    tabsize=1
}

\lstdefinestyle{mystyle}{
    backgroundcolor=\color{backcolour},   
    commentstyle=\color{codeblue},
    keywordstyle=\color{magenta},
    stringstyle=\color{codepurple},
    basicstyle=\ttfamily\footnotesize,
    breakatwhitespace=false,         
    breaklines=true,                 
    captionpos=b,                    
    keepspaces=true,                 
    numbers= none,                           
    showspaces=false,                
    showstringspaces=false,
    showtabs=false,                  
    tabsize=1
}

\begin{document}
%
\title{Identifying Technical Debt and Its Types Across Diverse Software Projects Issues}

\author{
    \IEEEauthorblockN{Karthik Shivashankar}
    \IEEEauthorblockA{University of Oslo\\
    karths@ifi.uio.no}
    \and
    \IEEEauthorblockN{Mili Orucevic}
    \IEEEauthorblockA{Visma Software International\\
    mili.orucevic@visma.com}
    \and
    \IEEEauthorblockN{Maren Maritsdatter Kruke}
    \IEEEauthorblockA{Visma Software International\\
    maren.kruke@visma.com}
    \and
    \IEEEauthorblockN{Antonio Martini}
    \IEEEauthorblockA{University of Oslo\\
    antonima@ifi.uio.no}
}


%


\maketitle

\begin{abstract}
Technical Debt (TD) identification in software projects issues is crucial for maintaining code quality, reducing long-term maintenance costs, and improving overall project health. This study advances TD classification using transformer-based models, addressing the critical need for accurate and efficient TD identification in large-scale software development.

Our methodology employs multiple binary classifiers for TD and its type, combined through ensemble learning, to enhance accuracy and robustness in detecting various forms of TD. We train and evaluate these models on a comprehensive dataset from GitHub Archive Issues (2015-2024), supplemented with industrial data validation.

We demonstrate that in-project fine-tuned transformer models significantly outperform task-specific fine-tuned models in TD classification, highlighting the importance of project-specific context in accurate TD identification.  Our research also reveals the superiority of specialized binary classifiers over multi-class models for TD and its type identification, enabling more targeted debt resolution strategies. A comparative analysis shows that the smaller DistilRoBERTa model is more effective than larger language models like GPTs for TD classification tasks, especially after fine-tuning, offering insights into efficient model selection for specific TD detection tasks. 

The study also assesses generalization capabilities using metrics such as MCC, AUC ROC, Recall, and F1 score, focusing on model effectiveness, fine-tuning impact, and relative performance. By validating our approach on out-of-distribution and real-world industrial datasets, we ensure practical applicability, addressing the diverse nature of software projects.

This research significantly enhances TD detection and offers a more nuanced understanding of TD types, contributing to improved software maintenance strategies in both academic and industrial settings. The release of our curated dataset aims to stimulate further advancements in TD classification research, ultimately enhancing software project outcomes and development practices by enabling early TD identification and management.
\end{abstract}


%
\IEEEpeerreviewmaketitle

\section{Introduction}

Technical Debt (TD) is the future cost of additional work in software development that arises from choosing quick, suboptimal solutions or from evolving requirements and technological advancements. Whether resulting from intentional decisions or external factors, TD necessitates careful management to maintain long-term project viability. Like financial debt, TD can accrue 'interest' over time, increasing complexity and maintenance costs  \cite{besker2020influence, holvitie2018technical}.

TD can be both intentional and unintentional. Intentional TD occurs when developers consciously opt for a faster solution to meet short-term goals, understanding that they will address the deficiencies later. Unintentional TD, on the other hand, results from inadvertent mistakes, poor practices, or a lack of knowledge or insights during the development process \cite{Fowler2009}. TD manifests in various forms, including complex and hard-to-maintain code, lack of documentation, duplicated code, and outdated libraries or frameworks. Such debts can degrade software quality, increase maintenance costs, and reduce developer productivity. Moreover, excessive TD can lead to missed business opportunities and potentially result in a development standstill, causing companies to lose clients. Consequently, effective management and timely resolution of TD are critical to maintaining the health, competitiveness, and sustainability of software projects  \cite{besker2020influence, Spinola2019}.

With the increasing software complexity and rapid development cycles, traditional methods of identifying and managing TD—such as manual code reviews and static analysis tools; often prove insufficient. These methods can be labour-intensive, prone to human error, and may not scale well with the size and complexity of modern software systems \cite{avgeriou2023technical}. This necessitates exploring automated techniques to detect and classify TD issues from project or source control management systems like Github or Jira, enabling timely interventions and proactive assessment of TD and its interest needed to be paid \cite{TDSkryseth}. Automated TD tracking offers several advantages. It reduces the workload associated with manual tracking in backlogs, which teams often resist due to its time-consuming nature. Additionally, it can help identify TD that might otherwise go unnoticed, educating team members on potential issues. Furthermore, it enables analytics on the quantity and types of TD accumulating, facilitating more informed decision-making and proactive management of technical debt \cite{ren2022identifying}. 

Recent advancements in Deep Learning, particularly the development of transformer-based models like BERT (Bidirectional Encoder Representations from Transformers) \cite{bert2018, roberta} and GPT (Generative Pre-trained Transformer) \cite{GPT2}, have opened new avenues for addressing complex natural language processing tasks with high accuracy. These models leverage deep learning techniques to understand and generate human-like text, making them well-suited for interpreting the often ambiguous and context-dependent descriptions in software documentation and issue trackers \cite{TransformerNLP}. Given these capabilities, applying transformer-based models to the domain of TD classification presents a promising approach. These models can potentially identify subtle nuances and patterns in text that traditional automated systems might miss. Furthermore, their ability to learn from vast amounts of unstructured data allows them to adapt to different programming languages and project specifics with minimal fine-tuning data \cite{TransformerNLP, maldonado2017natural}. 

Understanding and classifying different types of TD is crucial for effective management and mitigation strategies. However, accurately identifying and categorizing various TD types \cite{TDtypes} presents a significant challenge in software engineering research and practice.

Our study aims to explore these challenges by comparing different classification approaches and developing a ensemble method for TD and its type identification. We focus on the 13 types of TD identified in the ontology by Alves et al \cite{TDtypes}, including architectural, test, build, and code debt, among others. By leveraging a comprehensive dataset of GitHub issues from 2015 to 2024, we seek to develop and evaluate models that can accurately classify TD types with high precision and recall.

This research contributes to the field by addressing the complexities of TD classification and proposing solutions to enhance the accuracy and reliability of TD type identification in software projects.

A critical aspect of our research involves evaluating the models' ability to generalize to unseen data. This is essential for ensuring the practical applicability of our TD classification approach in real-world scenarios. To achieve this, we assess the models' performance on an out-of-distribution (OOD) dataset \cite{ODD1}, which includes projects not part of our training dataset.The use of OOD data provides a more rigorous test of our models' robustness and effectiveness in diverse, real-world software development environments.

\subsection{Research Questions}

This study seeks to answer the following key research questions:

\begin{itemize}
    \item \textbf{RQ1:} How effective are transformer-based models in classifying TD issues?
    \item \textbf{RQ1.1:} Does fine-tuning TD models on project-specific data improve their performance?
    \item \textbf{RQ2:} How does the performance of LLM-like the GPT model compare to the DistillRoberta models in TD classification?
    \item \textbf{RQ2.1:} Is task specific  fine-tuning of  LLM like GPT for TD   more effective than finetuned DistillRoberta TD classification?
    \item \textbf{RQ3:} How effective are expert ensemble of  binary classifiers in classifying different types of issues compared to  multi-class model?
    \item    \textbf{RQ3.1}: How does the performance of LLM-like the GPT model compare to the  DistillRoberta Model on different issues types?
\end{itemize}

Rationale for RQ1 and RQ1.1: Evaluating transformer models' effectiveness in TD classification is crucial for advancing automated detection. Also, to evaluate how fine-tuning may enhance performance by adapting to project-specific nuances and context.

Rationale for RQ2 and RQ2.1: Comparing GPT and DistillRoberta will provide insights into the balance between model size, GPU compute required, and classification performance, guiding future research and implementations.

Rationale for RQ3 and RQ3.1: Comparing an ensemble of multiple binary classifiers to a multi-class approach will determine the most effective strategy for handling diverse TD types. Assessing GPT and DistillRoberta across TD types will reveal their strengths in various aspects of technical debt and its type.

\subsection{Contributions}

This research makes several significant contributions to the field of software engineering:

\begin{itemize}
    \item \textbf{Empirical Assessment of Transformer-Based Models to classify TD:} This study comprehensively evaluates transformer-based models, specifically GPT and DistilRoBERTa, for TD classification. It examines the impact of fine-tuning on their performance, offering insights into their effectiveness and potential for practical application.
    \item \textbf{Specialised Binary Classifiers for TD Types:} We validate the use of multiple ensemble binary classifiers for identifying specific TD types (e.g., architectural, testing, build). This approach is compared with a multi-class classification approach, highlighting the advantages of using multiple ensemble binary models for precise and robust  TD type identification.
    \item \textbf{Generalisation to Out-of-Distribution Data:} The study empirically evaluates the models' ability to generalise to out-of-distribution (OOD) data, addressing a critical challenge for the practical adoption of TD classification tools in diverse software projects.
    \item \textbf{Publicly Available Dataset:} We will release our curated dataset of technical debt and  13 different types of TD  category issues sourced from GitHub Archive Issues events from 2015 to 2024. This dataset will support further research and development in TD classification, fostering advancements in software maintenance.
\end{itemize}

\section{Background}

\subsection{Transformer Models }

Transformer architectures have revolutionized the landscape of natural language processing (NLP), introducing a paradigm shift in how machines interpret and process human language. These models capture contextual relationships between words in a sentence, irrespective of their sequential order. At the heart of transformers lies the attention mechanism, which enables the model to weigh the importance of different input parts when producing an output \cite{NIPS2017_3f5ee243}. This approach allows transformers to establish global dependencies within the text, leading to more nuanced and context-aware language understanding 

Transformer models have a set performance benchmark in text classification, where the goal is to assign predefined categories to textual data. The Hugging Face (HF) ecosystem has democratized the implementation of these sophisticated models by providing a user-friendly interface and a comprehensive collection of pre-trained architectures. HF offers APIs for a wide range of transformer variants, including BERT \cite{bert2018}, GPT \cite{brown2020language}, and RoBERTa \cite{roberta}, enabling researchers to fine-tune these models on domain-specific datasets \cite{wolf2019huggingface}. This capability has revolutionized the field, allowing practitioners to achieve state-of-the-art results even with limited labelled data.

\subsection{Comparison of LLMs and BERT-based Models}

 Large Language Models (LLMs) like GPT-3 \cite{brown2020language} and GPT-4  have revolutionized NLP by leveraging vast amounts of training data to generate human-like text and perform a wide range of language tasks. These models are particularly known for their generative capabilities, making them versatile tools for various applications, including text classification. On the other hand, models like DistilRoBERTa, a distilled version of RoBERTa, offer a more efficient alternative with reduced computational requirements while retaining a significant portion of the original model's performance \cite{Sanh2019DistilBERTAD}.
 
 Several factors come into play when comparing the performance of LLMs like GPT 4 and DistilRoBERTa in TD classification. With their extensive training on diverse datasets, LLMs can handle complex language tasks and generate contextually relevant text. This makes them suitable for tasks that require a deep understanding of context and semantics. However, their large size and computational demands can be a drawback in resource-constrained environments \cite{roberta_vs_gpt4}.
 
  DistilRoBERTa, with its optimized architecture, offers a balance between performance and efficiency. It retains approximately 97\% of RoBERTa's performance while being faster and more resource-efficient. This makes it a viable option for TD classification tasks where computational resources are limited. Studies have shown that fine-tuning DistilRoBERTa on domain-specific data can enhance its performance, making it comparable to larger models like GPT-4 in certain tasks \cite{Sanh2019DistilBERTAD}.
 
 Both LLMs and DistilRoBERTa benefit from domain-specific training in terms of fine-tuning. Fine-tuning involves adapting the pre-trained model to the specific characteristics of the target domain, which can significantly improve classification accuracy \cite{finetune}. However, the choice between these models ultimately depends on the specific requirements of the task, including the available computational resources and the complexity of the classification problem. 

\section{Methodology}

The methodology for this research was designed to empirically evaluate the effectiveness of transformer-based models in classifying TD and its various types using issues extracted from GitHub. Our approach encompasses several stages: data mining, dataset processing and cleaning, model training, and model evaluation, including tests on out-of-distribution (OOD) datasets. Each stage is critical for ensuring the reliability and validity of the findings.

\subsection{Data Mining }

The initial stage of our methodology involved accumulating a large dataset of software development issues from GitHub, specifically targeting those related to TD and its 13 types. We utilized the GitHub Archive (GHArchive)  \cite{gharchive}, a project that records the public GitHub timeline, logs, and events. Data was mined from January 1, 2015, to May 25, 2024, to cover various software projects and incorporate various issue events.

In our study, we employed a carefully crafted regular expression (regex) pattern to identify TD related issues from GitHub Archive data. The regex, designed to be case-insensitive, targeted variations of the term "Technical debt" and its abbreviations in issue labels. Our pattern;
\begin{lstlisting}[breaklines=true]
r'(?i)\b(T(echnical[-_\s]?|ech[-_\s]?)?D(ebt|D)|\b(TD|td)\b|debt)\b'
\end{lstlisting}

effectively captured diverse representations such as "Technical debt", "Tech\_debt", "TD", and standalone "debt" mentions from the label field in the issue event from GHarchive. This approach ensured comprehensive coverage of TD-related discussions while minimizing false positives. The pattern's flexibility allowed for various word separators and abbreviations commonly used by developers, enhancing our data collection's accuracy and completeness. 
 
To identify issues related to specific types of TD \cite{TDtypes},  we employed a comprehensive regex pattern. The pattern was designed to capture 13  representations of TD types in a case-insensitive format within the 'labels' field of GitHub Issue events. Our regex pattern for issue types is as follows:
\begin{lstlisting}[breaklines=true]
r'(?i)\b(architect(ure|ural)?|build|code|defect|design|doc(umentation)?|infrastructure|people|process|requirement|service|test(ing)?|automation)\b'
\end{lstlisting}

The regex uses word boundaries (\textbackslash{}b) to ensure accurate matching and includes common variations (e.g., 'doc' for 'documentation', 'testing' for 'test'). This approach allows for a nuanced classification of TD types in software projects, enabling a more detailed analysis of TD distribution and impact. 

In addition to identifying issues related to TD and its specific types, we established a ground truth test dataset for our classification model evaluation. This dataset was created by selecting issues that explicitly contained both TD indicators with labels like "tech-debt" ,  "debt"  and "td"  and specific TD-type labels like "architecture", "build", "test" and other TD types \cite{TDtypes} that were unambiguously labelled by the project developers. These entries containing both these labels were removed from all our Datasets to avoid data leakage.This rigorous selection process provided a high-quality dataset for training and evaluating our TD classification models, enhancing the reliability and accuracy of our study results. 
 
\subsection{Dataset Processing and Cleaning}

Once the data was accumulated, it underwent extensive preprocessing and cleaning to prepare it for effective model training. The raw data from GHArchive \cite{gharchive} includes a lot of noise —duplicates, non-English text, irrelevant metadata, URLs, emojis, and special characters—that can obscure the underlying patterns related to TD. Therefore, the dataset was cleaned through several steps:
\begin{itemize}
    \item \textbf{Duplicate Removal}: Ensuring that each issue is unique to prevent bias in model training.
    \item \textbf{Text Normalization}: Converting all text to lowercase to maintain consistency across the dataset.
    \item \textbf{Noise Reduction}: We removed links, emojis, and special characters that did not contribute to understanding the context of TD.
    \item \textbf{Content Filtering}: Eliminating issues with very few characters likely does not contain enough classification information. 
\end{itemize}

Our study focused on 13 distinct categories of TD identified in the ontology by Alves et al: Architecture, Code, Defect, People, Automation, Build, Process, Design, Requirement, Infrastructure, Test, Service, and Documentation \cite{TDtypes} and TD itself. 

We created separate datasets for each category, with the number of instances varying significantly across categories, as shown in Table  \ref{tab:total_rows}.  To ensure balanced datasets for robust model training, we implemented a data cleaning process. For each category, we created equal proportions of positive and negative labels, dividing the total instances equally between the two classes. Sridharan et al. demonstrated the importance of data balance in SATD detection tasks, showing that imbalanced datasets can lead to biased models and unreliable predictions. This approach helps to mitigate the risk of model bias. It improves the overall reliability of our classification results across various TD types, extending the principles of balanced data from SATD detection to broader TD classification tasks.   \cite{sridharan2021data}. 

We created an out-of-distribution (OOD) dataset for robust evaluation for each category \cite{ODD1, ODD2}. We identified repositories with the highest number of issues for each type and withheld these from the training and testing datasets. This approach allowed us to assess the models' generalization capabilities across different project contexts. 

We then split each balanced dataset into training and testing sets using an 85/15 ratio. Despite the random nature of this split, we maintained the balance of positive and negative labels within each subset. This careful data preparation allowed us to develop more focused and accurate classification models for each TD type, enabling a nuanced analysis of TD in software development.

For the multiclass classification task, we adopted a different approach due to the inherent imbalance in TD type occurrences. We used only the positive labels from each category and employed stratified K-fold cross-validation for model training \cite{Kfold}. We maintained the 85/15 train-test split and also created an OOD dataset for the multiclass model to ensure a comprehensive evaluation.

This methodology enabled us to develop and rigorously test both binary and multiclass models for TD classification, providing a thorough analysis of TD detection and categorization in diverse software development scenarios.

\begin{table}[H]
\centering
\caption{Total Number of Instances for Different Categories}
\label{tab:total_rows}
\begin{tabular}{cc}
\begin{minipage}{0.3\linewidth}
\centering
\begin{tabular}{lr}
\toprule
      Category &  \#Instances \\
\midrule
Architecture &       20466 \\
        Code &      249934 \\
      Defect &      156356 \\
      People &        2020 \\
  Automation &       13356 \\
       Build &      191780 \\
     Process &       37532 \\
\bottomrule
\end{tabular}
\end{minipage} &
\begin{minipage}{0.4\linewidth}
\centering
\begin{tabular}{lr}
\toprule
      Category &  \#Instances \\
\midrule
        Design &      276320 \\
   Requirement &       26438 \\
Infrastructure &       59546 \\
          Test &      608634 \\
       Service &      130622 \\
 Documentation &      735926 \\
Technical Debt &      108653 \\
\bottomrule
\end{tabular}
\end{minipage}
\end{tabular}
\end{table}

\subsection{Model Training}

The training of the models was conducted using a carefully considered approach to ensure robustness and reliability. Binary classification models were trained on a balanced dataset to prevent any bias towards the more frequently occurring class \cite{sridharan2021data}. We employed 5-fold cross-validation over five epochs, a standard research practice that balances adequate model exposure to the data \cite{CV1, CV2}.

To achieve higher precision, recall and accuracy in classifying types of TD, we trained individual binary classifiers for each specific TD type (e.g., "test" vs "not test," "architectural" vs "not architectural," "code" vs. "not code"). This method employs an ensemble learning technique, combining multiple models to improve predictive performance by leveraging their strengths to enhance robustness and accuracy. Tan et al.  \cite{tan2021ensemble} demonstrated the effectiveness of ensemble learning techniques in improving predictive performance for Self-Admitted Technical Debt (SATD) detection tasks. They suggested that combining multiple models can enhance robustness and accuracy by leveraging the strengths of individual classifiers. Following their recommendations, we adopted an ensemble learning approach in our study for TD classification.  Our rule-based ensemble method integrates predictions from Classifier A and Classifier B. An instance is labelled as Architectural Technical Debt only if both classifiers produce positive results, indicating the presence of both architectural and technical debt.

For our multiclass classification models, we utilized Stratified 5-fold cross-validation over five epochs to address the imbalanced nature of different TD types' occurrences. This method ensures that each fold maintains the same proportion of classes as in the full dataset, providing a representative sample for model training and evaluation \cite{Kfold}. We incorporated class weights in computing the loss function to mitigate the impact of class imbalance in the multiclass classifier model. These weights were assigned to different classes based on their frequency in the dataset, giving more importance to underrepresented classes during training. This approach helps the model learn effectively from all classes, regardless of their prevalence in the dataset \cite{weighteclass}. By combining Stratified K-fold cross-validation with weighted loss computation, we aimed to develop robust multiclass models capable of accurately classifying various TD types despite their uneven distribution in real-world software projects.

Expert binary classifiers focus on identifying specific characteristics of their assigned type and enhance robustness and accuracy when combined with the general TD classifier. Furthermore, binary classifiers simplify the classification task to a binary decision, which is less sensitive to data imbalance than a MultiClass approach, ensuring accurate identification and management of even less frequently tagged TD types.

 \subsection{Model Evaluation and Testing with OOD Dataset}

The final stage of our methodology involved rigorous evaluation and testing of the trained models. We assessed model performance using standard metrics such as accuracy, precision, recall, Matthews’s correlation coefficient (MCC),  AUC (Area Under The Curve) ROC (Receiver Operating Characteristics), and F1-score on a reserved test set. 
Precision measures the percentage of TD issues that have been correctly identified (true positives) out of all predictions.
The Recall metric, which calculates the proportion of correctly identified TD issues (true positives) out of all actual TD issues (true positives + false negatives), holds particular importance in our study. 
Firstly, many software repositories contain only partially tagged TD issues. This means that the negative class in our dataset might inadvertently include some positive instances that were not explicitly labelled as TD. As a result, the precision metric, which relies on the accuracy of both positive and negative labels, cannot be considered entirely reliable in this scenario.In contrast, the Recall metric focuses solely on the correct identification of known TD issues, making it a more dependable measure of our model's performance. It assesses how effectively our approach captures all instances of TD, which is critical for comprehensive TD management.
 
The F1-score combines both precision and recall through a harmonic mean.
  The AUC is a performance measurement for classification problems at various threshold settings. It is the area under the Receiver Operating Characteristic (ROC) curve, which plots the true positive rate (TPR) against the false positive rate (FPR) at different threshold levels. The AUC value ranges from 0 to 1. An AUC of 0.5 suggests no discrimination (i.e., random guessing), while an AUC of 1 indicates perfect discrimination.

We also emphasize Matthews’s Correlation Coefficient (MCC), which is widely recognized as a more statistically reliable metric \cite{chicco_advantages_2020} for binary classification. 

{\scriptsize
\[
MCC = \frac{TP \cdot TN - FP \cdot FN}{\sqrt{(TP + FP) \cdot (TP + FN) \cdot (TN + FP) \cdot (TN + FN)}}
\]
}
MCC produces a more informative and truthful score in evaluating classifications than accuracy and F1 score. It produces a high score only if the prediction obtained good results in all four confusion matrix categories (true positives (TP), false negatives (FN), true negatives (TN), and false positives (FP)), proportionally both to the size of positive and negative elements in the dataset.
 MCC values range between 1 to -1, and a value greater than 0 indicates that the classifier is better than a random flip of a fair coin, whilst 0 means it is no better. 
 
In addition, all models were also tested against an out-of-distribution (OOD) dataset, which consisted of data from projects not included in the training dataset. This step is crucial to determining the models' generalization capabilities to new, unseen data contexts.

Furthermore, we also perform comparative analysis, including evaluating the performance of BERT-family models like DistilRoberta  \cite{Sanh2019DistilBERTAD} against the LLM like GPT \cite{brown2020language} Additionally, the effectiveness of binary classifiers for individual types of TD (such as architecture and testing) was compared against a multiclass approach where a single model attempted to classify all 13 different types of TD simultaneously.

Through this comprehensive methodology, we aimed to identify the most effective models and techniques for TD classification and contribute to a broader understanding of how Transformer-based models can be tailored and applied in software development and maintenance. This would provide actionable insights to help manage TD more effectively, leading to better-maintained and less costly software projects.

\section{Results}

We have made the replication package publicly available at  \textbf{Zenodo} https://zenodo.org/doi/10.5281/zenodo.12517204  .
This package includes all the code for training and evaluating the DistilRoberta and GPT models on various metrics. Additionally, we have made all datasets used in this paper available within this replication package except private project dataset from Visma. 

\subsection{RQ1: How effective are transformer-based models in classifying TD issues?}

 This research question aims to explore the effectiveness of the transformer-based model, particularly DistilRoberta in our case, in classifying TD issues, leveraging their ability to capture complex patterns and contextual information in text data. Understanding the effectiveness of these models can guide practitioners in selecting appropriate tools for automated TD management. This investigation will provide empirical evidence on the performance of transformer models in this specific domain, potentially leading to improved accuracy and efficiency in issues classification.

\begin{table}[htbp]
\centering
\caption{TD Performance metric on TestSet and OOD Dataset }
\label{tab:combined_performance_metrics}
\begin{tabular}{lccc}
\toprule
\textbf{Metric} & \textbf{TestSet}& \textbf{ OOD(va.gov-team)}& \textbf{OOD(VSCode)}\\
\midrule
Precision & 0.911 & 0.911 & 0.747 \\
Recall & 0.874 & 0.927 & 0.648 \\
Accuracy & 0.894 & 0.918 & 0.718 \\
MCC & 0.789 & 0.837 & 0.439 \\
F1 Score & 0.892 & 0.919 & 0.694 \\
AUC & 0.949 & 0.971 & 0.767 \\
\bottomrule
\end{tabular}
\end{table}

The results in Table \ref{tab:combined_performance_metrics} demonstrate Our TD (DistillRoberta)  binary classifier model's performance across different datasets. Notably, the model exhibits consistently high performance on both the testset and the out-of-distribution (OOD) va.gov-team dataset \cite{va_gov_team}, with precision, recall, and F1 scores all exceeding 0.90. An Out-Of-Distribution (OOD) dataset is used in machine learning to evaluate models' robustness and generalisation ability. OOD datasets contain samples that differ significantly from the data on which the model was trained. This helps to test how well the model can handle new, unseen data that does not fit the distribution of the training set. 
However, a performance drop is observed on the OOD VSCode dataset \cite{vscode}, particularly in the MCC metric, which indicates a decline in the model's ability to correctly classify positive instances. This highlights the challenge of domain shift, where the model's effectiveness diminishes when applied to data that diverges significantly from its training distribution.

These findings underscore the importance of considering domain specificity when deploying machine learning models and emphasize the need for further research into techniques like finetuning or transfer learning to enhance model robustness across diverse datasets.

\subsection{RQ1.1:Does fine-tuning TD models on project-specific data improve their performance?}

Fine-tuning models on domain-specific or project-specific datasets is a common practice to enhance performance. This research question seeks to determine whether fine-tuning improves the performance of our TD  (DistillRoberta ) model. The motivation is to assess the gains in classification accuracy and robustness when models are fine-tuned on project-specific or domain-specific TD data compared to their performance in our TD binary classifier. The results will inform how fine-tuning can be beneficial, providing actionable insights for researchers and practitioners looking to optimise their TD classification. 

For this study, we extract TD issues from the Jira Public dataset   \cite{JIRA_dataset}, which is an out-of-distribution (OOD) dataset, as the model was initially trained on GitHub issues.  We investigated the generalizability of this OOD Jira dataset to see how well the model can adapt to other project management tools before and after finetuning our TD model.

In addition, we also analysed the TD model's performance before and after fine-tuning with 30\% of VScode-specific TD data issues. The VScode project contains many TD issues and was removed from the initial TD dataset before training, ensuring the model has never encountered VScode-specific instances during training. 

We extended our investigation to examine the impact of fine-tuning on the model's temporal generalization capabilities. This aspect is crucial for understanding how well our TD classification models can adapt to and predict future TD issues.

To achieve this, we designed a temporal split in our dataset. We removed all TD-tagged data from GitHub issues dated 2024  (GH 2024) as shown in Table \ref{tab:performance_comparison_finetuning}  and used this as our test set. The model was then trained exclusively on historical TD and non-TD issues from before 2024. This approach simulates a real-world scenario where models are trained on past data and must predict future occurrences of TD.

By testing the model on the GH 2024 dataset, which it had never encountered during training, we aimed to evaluate its ability to generalize to future temporal data. This experiment allows us to assess how effectively our fine-tuning methods enable the model to capture underlying patterns of TD that persist over time, rather than merely memorizing specific instances or timebound characteristics.

\begin{table}[ht]
\centering
\caption{Comparison of  before and after fine-tuning (FT) TD model}
\begin{tabular}{lcccccc}
\toprule
\textbf{Model} & \textbf{Precision} & \textbf{Recall} & \textbf{Acc} & \textbf{F1} & \textbf{MCC} & \textbf{AUC} \\
\midrule
GH 2024 & 0.761 & 0.873 & 0.800 & 0.813 & 0.606 & 0.800 \\
30\% FT& 0.840 & 0.886 & 0.855 & 0.862 & 0.710 & 0.908 \\ \midrule
JIRA & 0.663 & 0.693 & 0.702 & 0.678 & 0.401 & 0.701 \\
30\% FT& 0.671 & 0.815 & 0.734 & 0.736 & 0.483 & 0.792 \\ \midrule
VSCode & 0.747 & 0.648 & 0.718 & 0.694 & 0.439 & 0.767 \\
30\% FT& 0.793 & 0.740 & 0.771 & 0.765 & 0.543 & 0.863 \\
\bottomrule
\end{tabular}

\label{tab:performance_comparison_finetuning}
\end{table}

 For this RQ, we have assessed improvements in classification accuracy, precision, recall, F1-score, MCC , and AUC ROC . The results, summarised in Table \ref{tab:performance_comparison_finetuning}, indicate significant performance gains through fine-tuning our TD (DistilRoBERTa) model.

Our analysis revealed that fine-tuned models using 30\% of project-specific data consistently outperformed their non-fine-tuned counterparts across all datasets. This improvement was evident in various key performance metrics.

For instance, when evaluating on the 'GH 2024' dataset split, which represents future data the model had not seen during training, the fine-tuned model demonstrated substantial improvements. Precision increased from 0.761 to 0.840, indicating a significant reduction in false positives. Recall also improved from 0.873 to 0.886, suggesting better identification of true TD issues. Overall accuracy saw a marked increase from 0.800 to 0.855, reflecting enhanced overall classification performance.Similar performance enhancements were observed across other datasets, including 'JIRA'  \cite{JIRA_dataset} and 'VSCode' \cite{vscode}.

 Skryseth et al. conducted an empirical evaluation to determine the optimal percentage of project-specific data for fine-tuning TD classification models. Their study examined a range of percentages from 15\% to 50\% and found that 30\% of project-specific data consistently yielded good results across various scenarios \cite{TDSkryseth}.  
 Following their findings, we chose to use 30\% of project-specific data for fine-tuning our models. This percentage strikes a balance between leveraging project-specific information and maintaining model generalizability. It provides sufficient data to capture project-specific TD patterns without overfitting to individual project idiosyncrasies \cite{TDSkryseth}. However, it's important to note that practitioners can adjust this percentage based on their specific circumstances and data availability. The 30\% guideline serves as a well-supported starting point, but the optimal percentage may vary depending on factors such as project size, TD prevalence, and the diversity of TD types within a project. These findings underscore the effectiveness of fine-tuning in enhancing model robustness and accuracy, providing valuable insights for researchers and practitioners aiming to optimise TD classification frameworks.

\subsection{RQ2: How does the performance of LLM-like the GPT model, compare to TD DistilRoberta (Our) models in TD classification?}

Large Language Models (LLMs) like GPT and smaller, efficient models like DistilRoberta represent different trade-offs in NLP tasks. This research question aims to compare the performance of these two models in the context of TD classification. The motivation is to understand the trade-offs between the computational efficiency of DistilRoberta and the expansive capabilities of GPT. The comparison will help identify the optimal model for practical applications, balancing accuracy, computational cost, and resource requirements. The result will clearly explain which model is more suitable for TD classification tasks.

\begin{table}[ht]
\centering
\caption{Comparison of GPT4o and TD DistilRoberta (Our) Model Performance Metrics}
\label{tab: GPT4o and DistilRoberta}
\begin{tabular}{lccccccc}
\toprule
\textbf{Metric} & \textbf{GPT4o} & \textbf{Our TD model} & \textbf{Improvement} \\
\midrule
Precision & 0.634 & 0.761 & +20.03\% \\
Recall & 0.630 & 0.873 & +38.57\% \\
Accuracy & 0.633 & 0.800 & +26.38\% \\
F1-Score & 0.632 & 0.813 & +28.64\% \\
MCC & 0.267 & 0.606 & +126.97\% \\
AUC & 0.633 & 0.800 & +26.38\% \\
\bottomrule
\end{tabular}
\end{table}

 Our study compared the performance of our fine-tuned TD DistilRoberta model against the out-of-the-box GPT4o (“o” for “omni”) 
  model on the same test set, revealing significant improvements across all metrics as shown in Table \ref{tab: GPT4o and DistilRoberta} .

DistilRoberta outperformed GPT4o in precision (0.761 vs 0.634, a 20.03\% improvement), recall (0.873 vs 0.630, a 38.57\% increase), and accuracy (0.800 vs 0.633, a 26.38\% gain). The F1-Score for DistilRoberta was 0.813, 28.64\% higher than GPT4o's 0.632. Notably, the MCC showed the most substantial difference, with DistilRoberta scoring 0.606 compared to GPT4o's 0.267, a 126.97\% increase. The AUC for DistilRoberta (0.800) also surpassed GPT4o (0.633) by 26.38\%.

These results demonstrate the superior performance of our fine-tuned DistilRoberta model across all key metrics. This comparison underscores the effectiveness of fine-tuning in enhancing model performance and highlights the potential of the smaller, more efficient DistilRoberta model in achieving high accuracy in TD classification tasks.

\subsection{RQ2.1: Is task specific  fine-tuning of  LLM like GPT for TD   more effective than finetuned DistillRoberta TD classification?}

In this RQ, we compares the performance of task-specific fine-tuned GPT models (GPT-3.5 Turbo and GPT-DaVinci-002) against our task-specific fine-tuned TD DistilRoBERTa model for Technical Debt classification on the same Train and test split. We distinguish between two types of fine-tuning:

\begin{enumerate}
    \item Task-specific fine-tuning: Adapting pre-trained models to the TD classification task.
    \item Project-specific fine-tuning: Further adapting models to specific project data.
\end{enumerate}

Our focus is on task-specific fine-tuning, examining whether the extensive pre-training of larger GPT models yields superior performance compared to the smaller, more efficient DistilRoBERTa  \cite{roberta, Sanh2019DistilBERTAD} when both are fine-tuned for TD classification.

We consider the trade-offs between computational resources and model efficiency. GPT models, while powerful, are closed-source and costly for fine-tuning and inference. In contrast, DistilRoBERTa (87 million parameters) can be fine-tuned on smaller GPUs or free resources like Google Colab, and converted to ONNX format for CPU inference \cite{onnx}.

This investigation aims to guide practitioners in model selection and resource allocation for TD classification tasks, balancing performance gains against computational costs and accessibility.

\begin{table}[h!]
\centering
\caption{Comparing the GPT with DistilRoberta (ours) FT model}
\label{tab:Comparing GPT's with DistilRoberta FT}
\begin{tabular}{lcccc}
\toprule
Metric & GPT-3.5 & Davinci & Ours & Ours VS GPT 3.5(\%) \\
\midrule
Precision (\%)& 80.6 & 62.1 & 91.1 & 13.0 \\
Recall (\%) & 89.9 & 98.3 & 87.4 & -2.8 \\
Accuracy (\%) & 84.1 & 69.2 & 89.4 & 6.3 \\
F1-Score (\%) & 85.0 & 76.1 & 89.2 & 4.9 \\
MCC (\%) & 68.7 & 47.2 & 78.9 & 14.9 \\
AUC (\%) & 84.1 & 69.2 & 94.9 & 12.8 \\
\bottomrule
\end{tabular}
\end{table}

We compared the performance of task-specific fine-tuned GPT-3.5 turbo, GPT-DaVinci-002, and our TD DistilRoBERTa model for Technical Debt classification. All models were trained and tested on the same datasets to ensure a fair comparison. The results revealed that our TD DistilRoBERTa model generally outperformed the larger GPT models across most metrics as show in Table \ref{tab:Comparing GPT's with DistilRoberta FT}  .

DistilRoBERTa demonstrated superior performance in precision, accuracy, F1-Score, and MCC. It achieved the highest precision at 91.1\%, significantly surpassing both GPT models. In terms of overall accuracy and F1-Score, DistilRoBERTa also led, indicating a well-balanced and reliable performance. The MCC, which provides a comprehensive measure of classification quality, was notably higher for DistilRoBERTa, further confirming its robust performance.

While GPT-DaVinci-002 showed the highest recall, DistilRoBERTa's performance in this metric was competitive and close to that of GPT-3.5 turbo. These results suggest that despite the computational advantages often associated with larger models, a smaller, efficiently fine-tuned model like DistilRoBERTa can achieve superior performance in specialized tasks such as TD classification. This finding emphasizes the potential cost-benefit advantage of using more resource-efficient models without significant performance loss in practical applications.

\subsection{RQ3:  How effective are expert ensemble of binary classifiers in classifying different types of issues compared to multi-class model?}

\subsubsection{Evaluation binary Classifier }

This research question investigates the effectiveness of classification models in distinguishing among 13 different types of software issues, as identified by Alves et al \cite{TDtypes}. These issue types are not explicitly labelled as TD but represent categories often associated with TD. For the TD identification, we employ a separate expert binary TD model.

Our primary goal is to improve the granularity of issue identification, which could lead to more targeted remediation strategies, whether these issues represent TD or not. 
We compare two approaches:

\begin{enumerate}
    \item A multiclass classification model that attempts to categorize issues into all 13 types simultaneously.
    \item Expert binary classifiers, each trained to identify a specific issue type (e.g., "architectural" vs. "not architectural").
\end{enumerate}

The dataset for this study was curated by mining GitHub issues labelled with keywords corresponding to various tissue types, such as Architecture, Automation, Build, and Code. We divided this data into training, test, and separate out-of-distribution (OOD) projects not included in the training phase.

We assess the performance of both multiclass and binary models on the test set and the OOD dataset. This evaluation aims to determine which approach - multiclass or multiple expert binary classifiers - is more effective in distinguishing between different issue types.

\begin{table}[ht]
\caption{Test Set and OOD Performance Metrics for Various Categories by Binary classifiers }
\centering
\begin{tabular}{c ccc ccc}
\toprule
\textbf{Category} & \multicolumn{3}{c}{\textbf{Test Set}} & \multicolumn{3}{c}{\textbf{OOD}} \\
\cmidrule(lr){2-4} \cmidrule(lr){5-7}
& \textbf{MCC} & \textbf{F1} & \textbf{Recall}& \textbf{MCC} & \textbf{F1} & \textbf{Recall}\\
\midrule
Architecture & 0.616 & 0.809 & 0.817 
& 0.803 & 0.903 & 0.992 
\\
Automation & 0.680 & 0.836 & 0.819 
& -0.124 & 0.096 & 0.057 
\\
Build & 0.760 & 0.880 & 0.878 
& 0.836 & 0.920 & 0.957 
\\
Code & 0.592 & 0.793 & 0.780 
& -0.027 & 0.246 & 0.166 
\\
Defect & 0.865 & 0.932 & 0.925 
& 0.860 & 0.930 & 0.925 
\\
Design & 0.692 & 0.847 & 0.854 
& 0.780 & 0.893 & 0.939 
\\
Documentation & 0.773 & 0.887 & 0.889 
& 0.719 & 0.857 & 0.846 
\\
Infrastructure & 0.642 & 0.820 & 0.815 
& 0.743 & 0.876 & 0.913 
\\
People & 0.769 & 0.887 & 0.907 
& 0.887 & 0.944 & 1.000 
\\
Process & 0.605 & 0.800 & 0.792 
& -0.010 & 0.250 & 0.168 
\\
Requirement & 0.678 & 0.843 & 0.865 
& 0.774 & 0.891 & 0.951 
\\
Service & 0.813 & 0.906 & 0.897 
& 0.892 & 0.947 & 0.974 
\\
Test & 0.731 & 0.864 & 0.851 & 0.862 & 0.932 & 0.977 \\
\bottomrule
\end{tabular}
\label{tab:test_cat_performance_metrics}
\end{table}

Table \ref{tab:test_cat_performance_metrics}  presents performance metrics for various issue categories on both the test set and out-of-distribution (OOD) data, focusing on MCC, F1 score, and Recall using DistilRoBERTa binary classifier.

On the test set, most categories show strong performance with MCC values typically above 0.6 and F1 scores ranging from 0.8 to 0.9. Categories like "Defect," "Service," and "Build" perform exceptionally well, with MCC values around or above 0.8.

However, OOD data reveals significant performance variations. Categories such as "Architecture," "Build," "Defect," "Service," and "People" maintain high MCC values, F1 scores, and Recall, indicating good generalization. In contrast, "Automation," "Code," and "Process" show sharp declines in all metrics, with negative MCC values suggesting poor performance on OOD data.

These results highlight the variability in model generalization across different issue categories. While some models demonstrate robust performance beyond their training data, others struggle significantly with OOD samples. This disparity underscores the need for targeted improvements in specific categories to enhance model robustness and applicability in diverse, real-world scenarios.

 \subsubsection{Evaluation and comparison of  Multi-class classifier}

We evaluate the effectiveness of multi-class classifiers in distinguishing between 13 different types of software issues often associated with TD, such as Architecture, Code, and Build. It's important to note that the multi-class model is trained to identify issue categories, not the presence of TD itself. We compare this multi-class approach with the binary classifiers. The motivation is to understand the potential of multi-class models in providing comprehensive and accurate issue-type identification.

This comparison aims to identify the relative strengths and weaknesses of each approach. Multi-class classification offers a holistic view but may face challenges in complexity and computational costs. Binary classifiers, while simpler and potentially more interpretable, might be less comprehensive when considering all issue types simultaneously.
The findings will contribute to developing more sophisticated classification systems and improving software quality assessment and maintainability strategies.

\begin{table}[h]
\caption{Performance Metrics for Multi-Class (MC) and Binary Models on various categories }
\centering
\begin{tabular}{c c c c c c c}
\toprule
\textbf{Metric} & \textbf{MC Test} & \textbf{MC OOD} & \multicolumn{2}{c}{\textbf{Binary Test}} & \multicolumn{2}{c}{\textbf{Binary OOD}} \\
\cmidrule(lr){4-5} \cmidrule(lr){6-7}
 & & & \textbf{Mean} & \textbf{SD} & \textbf{Mean} & \textbf{SD} \\
\midrule
Precision & 0.666 & 0.588 & 0.855 & 0.043 & 0.771 & 0.205 \\
Recall & 0.648 & 0.593 & 0.853 & 0.046 & 0.759 & 0.361 \\
Accuracy & 0.761 & 0.627 & 0.854 & 0.042 & 0.808 & 0.187 \\
MCC & 0.710 & 0.561 & 0.709 & 0.084 & 0.615 & 0.386 \\
F1 Score & 0.655 & 0.554 & 0.854 & 0.043 & 0.745 & 0.315 \\
AUC & 0.929 & 0.859 & 0.923 & 0.030 & 0.865 & 0.202 \\
\bottomrule
\end{tabular}
\label{tab:combined_performance_metrics_both}
\end{table}

Table \ref{tab:combined_performance_metrics_both} shows the multi-class (MC) DistilRoBERTa classifier's performance, with test set metrics of Precision (0.666), Recall (0.648), Accuracy (0.761), MCC (0.710), F1 Score (0.655), and AUC (0.929). Out-of-distribution (OOD) results are slightly lower, indicating some challenges in generalization.
Table \ref{tab:combined_performance_metrics_both}  also summarizes the performance of binary DistilRoBERTa classifiers for 13 TD types from results in Table \ref{tab:test_cat_performance_metrics} with mean and Standard Deviation (SD). On the test set, mean metrics include Precision (0.855), Recall (0.853), Accuracy (0.854), MCC (0.709), F1 Score (0.854), and AUC (0.923). OOD performance shows more variability but remains strong.

Comparing the two approaches:
\begin{itemize}

    \item Binary classifiers demonstrate higher precision and recall in both test and OOD datasets.
    \item Binary models achieve higher accuracy, suggesting more reliable distinction between TD types.
    \item Binary classifiers outperform the multi-class model in MCC and F1 score.
\end{itemize}

Overall, binary classifiers exhibit superior performance metrics. This suggests that combining an ensemble of binary classifier categorization with expert binary TD models could leverage the strengths of both approaches, potentially enhancing overall TD identification and management.

\subsection{RQ 3.1:  How does the performance of LLM-like the GPT model compare to the  DistillRoberta (ours) Model on different issue types ?}

Table \ref{tab:performance_comparison_2024_Our_GPT4o} compares performance metrics between our fine-tuned DistilRoberta model and GPT-4o (“\textbf{o}” for “omni”)  across various issues categories.
\begin{table}[H]
    \centering
    \caption{Comparison of Performance Metrics between Ours and GPT-4o}
    \begin{tabular}{lccc|ccc}
        \toprule
        Category & \multicolumn{3}{c}{Ours} & \multicolumn{3}{c}{GPT-4o} \\
                 & F1-Score & MCC & Recall& F1-Score & MCC & Recall\\
        \midrule
        Arch
& 0.803 
& 0.622 & 0.773 
& 0.758 & 0.543 & 0.720 
\\
        Auto
& 0.766 
& 0.582 & 0.700 
& 0.759 & 0.551 & 0.713 
\\
        Build 
& 0.881 
& 0.767 & 0.867 
& 0.875 & 0.753 & 0.867 
\\
        Code 
& 0.703 
& 0.459 & 0.647 
& 0.675 & 0.165 & 0.920 
\\
        Defect 
& 0.921 
& 0.848 & 0.893 
& 0.760 & 0.466 & 0.980 
\\
        Design 
& 0.904 
& 0.814 & 0.880 
& 0.809 & 0.645 & 0.760 
\\
        Doc
& 0.886 
& 0.768 & 0.907 
& 0.868 & 0.749 & 0.833 
\\
        Infra
& 0.769 
& 0.562 & 0.733 
& 0.719 & 0.370 & 0.827 
\\
        People 
& 0.944 
& 0.889 & 0.944 
& 0.971 & 0.946 & 0.944 
\\
        Process 
& 0.751 
& 0.566 & 0.673 
& 0.714 & 0.412 & 0.738 
\\
        Require
& 0.836 
& 0.681 & 0.813 
& 0.785 & 0.582 & 0.787\\
        Service 
& 0.861 
& 0.746 & 0.807 
& 0.800 & 0.572 & 0.880 
\\
        Test & 0.897 & 0.793 & 0.900 & 0.794 & 0.618 & 
0.747\\
        \bottomrule
    \end{tabular}
    \label{tab:performance_comparison_2024_Our_GPT4o}
\end{table}
We compare the performance of our fine-tuned DistilRoberta model with GPT-4o across various issue categories commonly associated with software development and potential Technical Debt types. Both models were evaluated on a 2024 dataset to ensure fair comparison and avoid data leakage, as neither model was trained on this data.
Our DistilRoberta model demonstrated superior performance in most technical categories. For example, in 'Defect' detection, it achieved an F1-Score of 0.923, significantly outperforming GPT-4's 0.760. This trend was consistent across several technical categories, indicating our model's strength in areas requiring high technical accuracy and automation.
However, GPT-4 excelled in the 'People' category, with an F1-Score of 0.971 compared to our model's 0.944. 
These results highlight the effectiveness of our fine-tuned DistilRoberta model in classifying all TD issue types. 

\section{Evaluation}
In our evaluation study, we focused on two main classification tasks:
\begin{enumerate}
    \item Categorizing software issues into 13 types commonly associated with TD, such as architecture, build, and test.
    \item Classifying and identifying whether an issue represents TD or Not using an expert TD Binary model.
\end{enumerate}
We created a ground truth dataset consisting of issues tagged with both a specific category label and a TD label. For example, an issue might be labelled as both "Architecture" and "Technical Debt."
To ensure a fair evaluation, we excluded these doubly-tagged issues from our training data and used them as a separate test set. We then evaluated three types of models:
\begin{enumerate}
    \item A multi-class model that categorizes issues into all 13 types simultaneously.
    \item 13 separate binary models, each specialized in identifying one specific issue type.
    \item A single expert binary model focused solely on identifying TD.
\end{enumerate}
This approach allowed us to compare the effectiveness of different classification strategies and assess how well each model type performs in identifying specific issue categories and TD.

We used the recall metric to evaluate model performance. Recall was chosen because our ground truth dataset only contains positive examples (issues tagged with both a specific TD category and TD). We don't have negative examples of doubly-tagged issues in our ground truth. Recall measures how many of these known positive instances were correctly identified by each model.
For instance, if we have 100 issues tagged as both "Architecture" and "TD," and our model correctly identifies 80 of them, the recall would be 0.8 or 80\%. 
The support metric indicates the total number of these doubly-tagged issues for each category, providing context for the recall scores.
This approach allowed us to assess how well our models could identify specific issue categories and TD, using a carefully curated set of real-world, developer-labeled issues as our benchmark.

\begin{table}[h]
\centering
\caption{OOD Test TD Types Classification of our DistilRoBERTa}
\begin{tabular}{lcccc}
\toprule
\multirow{2}{*}{\textbf{Category}} & \multicolumn{3}{c}{\textbf{Recall}} & \multirow{2}{*}{\textbf{Support}} \\
\cmidrule(lr){2-4}
 & \textbf{MC} & \textbf{Binary types}& \textbf{TD only }& \\
\midrule
Architecture Debt & 0.160 & 0.755 & 0.980 & 49 \\
Automation Debt & 0.470 & 0.975 & 0.980 & 40 \\
Build Debt & 0.600 & 0.829 & 0.947 & 246 \\
Code Debt & 0.610 & 0.821 & 0.918 & 469 \\
Defect Debt & 0.740 & 0.800 & 0.980 & 50 \\
Design Debt & 0.650 & 0.800 & 0.975 & 670 \\
Documentation Debt & 0.790 & 0.803 & 0.967 & 528 \\
Infrastructure Debt & 0.300 & 0.774 & 0.954 & 195 \\
People Debt & 0.000 & 0.856 & 0.930 & 1 \\
Process Debt & 0.210 & 0.727 & 0.939 & 66 \\
Requirement Debt & 0.230 & 0.736 & 0.970 & 61 \\
Service Debt & 0.670 & 0.790 & 0.984 & 433 \\
Test Debt & 0.670 & 0.825 & 0.972 & 1950 \\
\bottomrule
\end{tabular}
\label{tab:OOD_Test_TD_types_MC_and_Binary}
\end{table}

Table \ref{tab:OOD_Test_TD_types_MC_and_Binary} compares the performance of our DistilRoBERTa multi-class (MC), binary issues categories type, and expert TD only classifiers in identifying various types of technical debt. The results show that binary classifiers generally outperform the multi-class approach in terms of recall across most TD types.
Key findings:

\begin{itemize}
    \item Binary classifiers consistently achieve higher recall than the multi-class (MC)  model. For example, in Architecture Debt detection, recall values are 0.160 (MC), 0.755 (binary architecture issues type), and 0.980 (expert TD only).
    \item The expert TD binary model demonstrates consistently high recall across all categories, highlighting the effectiveness of specialized models.
    \item Support values vary significantly across categories, influencing the reliability of recall metrics. High-support categories like Test Debt (1950 instances) offer more robust results, while low-support categories like People Debt (1 instance) may yield less reliable metrics.
\end{itemize}

These results suggest that an ensemble of binary classification, with specialized models for each TD type and expert TD-only model, is more effective than a multi-class approach for technical debt identification.

\subsection{Evaluation of Industry projects}
We evaluated the DistillRoBERTa TD binary classification model on a dataset from Visma, a multinational software company comprising over 180 entrepreneurial software companies. Visma's unique structure, with high autonomy among its companies and diverse team practices, makes it an ideal candidate for testing the model's effectiveness in identifying and classifying TD across varied project contexts.
This dataset from Visma consisted of approximately 1100 Jira issues from 6 different projects across 3 countries, with about 400 issues labelled as TD. This diversity in data sources ensures a robust test of the model's generalization capabilities.
Table \ref{tab:company_classification_metrics_combined} presents our DistilRoBERTa model's performance before and after fine-tuning with 30\% of the Visma dataset:
 
\begin{table}[h]
    \centering
    \caption{ Classification Metrics for Technical Debt Before and After 30\% Fine-Tuning with Visma Dataset}
    \begin{tabular}{c c c}
        \toprule
        \textbf{Metric} & \textbf{Before Fine-Tuning} & \textbf{After 30\% Fine-Tuning} \\ \midrule

        Accuracy & 0.568 & 0.737 \\ 
        Precision & 0.451 & 0.608 \\ 
        Recall & 0.791 & 0.801 \\ 
        F1-Score & 0.574 & 0.691 \\ 
        MCC & 0.664 & 0.707 \\ 
        AUC & 0.615 & 0.750 \\ \bottomrule
    \end{tabular}
    
    \label{tab:company_classification_metrics_combined}
\end{table}
The results show significant improvements across all metrics after fine-tuning. Notably, accuracy increased from 56.8\% to 73.7\%, and precision from 45.1\% to 60.8\%. The recall, crucial for identifying potential TD issues, was already high at 79.1\% and improved slightly to 80.1\%.
In this context, recall is particularly important. Given the inconsistent TD tagging practices across different teams and projects, the model's ability to identify potential TD issues, even those not explicitly tagged, is vital. A high recall ensures that most TD issues are captured, allowing teams to address them proactively.
The improvements in F1-Score (57.4\% to 69.1\%), MCC (66.4\% to 70.7\%), and AUC (61.5\% to 75.0\%) further demonstrate the model's enhanced effectiveness after fine-tuning.
These results are promising for Visma's diverse software ecosystem. The model's improved performance in identifying TD issues, even in previously untagged instances, can significantly aid in standardizing TD management practices across Visma's various companies. This approach enhances software maintenance practices and contributes to more sustainable and consistent software development processes within Visma's complex corporate environment.

\section{Discussion and Related Works}
\subsection{The Importance of Recall in TD Detection}
In TD classification, recall plays a crucial role. High recall ensures that most TD instances are identified, minimizing the risk of overlooking critical issues. Our models achieved a balance between precision and recall, with a slight emphasis on recall. This approach is particularly valuable in large-scale projects where manual review of all potential TD is impractical.
For example, in a large enterprise software project with thousands of issues, our model might identify 90 to 80\% of actual TD instances (high recall) while maintaining a manageable number of false positives. This allows project managers to focus on addressing the most critical TD issues without being overwhelmed by false alarms.

\subsection{Implications for Research}
\subsubsection{Large-scale TD Studies}
Our models enable reliable tagging of large numbers of issues, facilitating comprehensive studies on TD patterns and evolution. For instance, researchers could use our models to analyze TD trends across thousands of open-source projects on GitHub, providing insights into common TD types and their prevalence across different programming languages or project domains \cite{bavota2016large}.
\subsubsection{Temporal Analysis of TD}
The ability of our models to generalize to future data opens possibilities for studying TD evolution over time. Researchers could track how TD accumulates in long-running projects, potentially developing predictive models for TD growth based on project characteristics and development practices.
\subsubsection{Cross-project TD Analysis}
Our approach's effectiveness across different projects enables comparative studies of TD across diverse software ecosystems. For example, researchers could compare TD patterns in mobile app development versus enterprise software projects, identifying domain-specific TD challenges.
\subsection{Implications for Industry}
Our research offers several practical benefits for the software industry:
\subsubsection{Retroactive TD Tagging}
Our approach allows for retroactively tagging TD in existing projects. This is valuable for scenarios such as \cite{ernst2021technical}:
\begin{itemize}
    \item Postmortem analysis of completed projects to identify areas for improvement in future developments.
    \item Due diligence during company acquisitions, providing insight into the acquired software's technical health.
    \item Project handovers, ensuring the new team understands the existing TD landscape.
\end{itemize}
\subsubsection{Efficient TD Management}
The deployment of our models can streamline TD detection and management in large projects. For instance, a development team could integrate our model into their CI/CD pipeline, automatically flagging potential TD issues during code reviews and prioritizing them based on severity \cite{TDSkryseth}.
\subsubsection{Customizable TD Detection}
The ability to fine-tune models on project-specific data allows companies to tailor TD detection to their contexts. A financial services company, for example, could fine-tune the model to be particularly sensitive to security-related TD, reflecting their industry's specific concerns.
\subsection{Integration with Software Economics and Project Management}
Our TD classification approach has broader implications:
\subsubsection{TD Quantification}
Future research could develop methods to quantify the economic impact of identified TD. For example, by estimating the cost of addressing TD versus the long-term savings from reduced maintenance efforts, organizations could make more informed decisions about TD remediation priorities \cite{lenarduzzi2021systematic, tsoukalas2018methods, arvanitou2022quantifying}.
\subsubsection{TD in Agile Methodologies}
Investigating how our automated TD classification can be integrated into agile processes could provide insights into balancing short-term delivery with long-term code quality. For instance, TD metrics could be incorporated into sprint planning and retrospectives, ensuring TD is consistently addressed alongside new feature development \cite{santos2022technical}.
\subsubsection{TD and Software Metrics}
Exploring correlations between our TD classifications and traditional software metrics could provide a more holistic view of software quality. Researchers could investigate how TD correlates with metrics like cyclomatic complexity or coupling, potentially developing more comprehensive software quality models \cite{tsoukalas2018methods, arvanitou2022quantifying}.

\subsection{Limitations and Future Work}
Our study, while promising, has limitations. The keyword-based dataset curation might miss implicit TD instances, suggesting future exploration of advanced NLP techniques for comprehensive TD identification.

While we evaluated models on OOD datasets, further testing on diverse real-world projects is needed to confirm robustness and generalizability. Future work should involve live deployments and developer feedback to refine the system \cite{avgeriou2023technical}.

Although our ensemble learning approach with binary classifiers showed promising results, there's room for optimization. Future research could explore advanced ensemble techniques like stacking or boosting to enhance predictive performance.

Expanding the dataset to include a wider range of sources and project types would ensure broader model applicability, addressing the current limitations in project diversity representation.

\subsection{Related Works}
The impact of Technical Debt (TD) on software quality and maintenance costs has led researchers to explore increasingly sophisticated automated detection and management methods. One prominent approach focuses on identifying self-admitted technical debt (SATD) in source code comments. Maldonado et al. utilized NLP techniques to automatically detect SATD, achieving high precision and recall in identifying design and requirement debt \cite{maldonado2017natural}. Building on this foundation, Ren et al. developed a convolutional neural network (CNN) based approach, which outperformed traditional machine learning methods in accuracy and F1-score \cite{ren2019neural}.
To enhance the interpretability of SATD detection models, Huang et al. proposed a graph neural network (GNN) based approach, providing not only high accuracy but also explanations for predictions \cite{huang2022exploiting}. Recognizing the unique manifestations of TD in machine learning (ML) systems, Sculley et al. introduced the concept of "hidden technical debt in machine learning systems," identifying ML-specific debt types and their implications \cite{sculley2015hidden}. Guo et al. further addressed these challenges through an empirical study on refactoring and TD in ML systems, offering insights into the unique characteristics of TD in ML code \cite{guo2021empirical}.
Recent advancements in TD detection include DebtHunter by Liu et al., a machine learning-based approach combining code metrics and textual features with ensemble learning techniques to enhance detection accuracy \cite{liu2021debthunter}. Huang et al. developed SATD Detector, a text-mining-based tool using NLP techniques and machine learning classifiers to identify SATD in source code comments \cite{huang2018satd}. Ren et al. explored SATD detection in issue tracking systems using various text classification techniques, and to tackle limited labeled data, they proposed DebtFree, a semi-supervised learning approach to minimize labeling costs while maintaining high accuracy \cite{ren2022identifying, ren2022debtfree}.
In our previous study \cite{TDSkryseth}, we demonstrated that transformer-based models, like DeBERTa \cite{he2021deberta}, exhibit exceptional performance in TD classification tasks. Building on this foundation, our current work significantly extends the scope and applicability of TD detection. We now classify 13 distinct TD types, offering a more comprehensive understanding of TD in software projects. Our study is the first to comprehensively compare multi-class and binary classification approaches for TD types, as well as evaluate the performance of transformer models like DistilRoBERTa and GPTs like GPT 3.5 turbo and 4o in this context.
Furthermore, Flisar and Podgorelec investigated deep learning methods for SATD detection, comparing various neural network architectures to traditional machine learning approaches \cite{flisar2021deep}. Tan et al. proposed an ensemble learning approach, combining multiple classifiers to improve SATD detection accuracy and robustness \cite{tan2021ensemble}.
Our current work not only builds upon these existing studies but also significantly advances the field by introducing novel contributions. We validate our models on a real-world dataset from a multinational software company, demonstrating their practical applicability and generalizability.

\section{Threats to validity}

In our study, we identify threats to validity that could impact the results and interpretations.

A key threat to internal validity is the potential for selection bias in our dataset. Our dataset comprises issues tagged by developers with terms related to technical debt, which may not represent all instances of TD comprehensively. Additionally, the accuracy of our classifiers might be influenced by the quality of the annotations. Any inconsistencies or inaccuracies in tagging could affect the model's performance. Moreover, the use of transformer-based models, which require substantial computational resources, might introduce variance in results due to differences in training conditions and hyperparameter settings.

The generalisability of our findings is a significant concern. Our study uses a dataset curated from GitHub projects, which might not represent all types of software projects or development environments. This limitation could affect the applicability of our results to other domains or projects with different characteristics. Additionally, the models' performance on out-of-distribution (OOD) data needs careful interpretation, as it may not accurately reflect real-world scenarios where data characteristics differ significantly from the training set.
However, it's important to note that we have taken steps to address this concern. We evaluated our models on multiple OOD projects and, notably, on industrial projects from a multinational software company. These evaluations yielded promising results, suggesting a degree of generalisability beyond our initial dataset.

Construct validity threats include the definitions and measurements of technical debt types. While we rely on established ontologies, the subjective nature of tagging and the varying interpretations of what constitutes technical debt could introduce bias. Furthermore, our approach to classifying TD types using binary classifiers assumes clear distinctions between types, which may not always hold true in practice. The overlap between different types of technical debt could affect the accuracy of our classifications and the conclusions drawn from them.
The potential overlap between different types of technical debt could affect the granularity of our classifications to some degree. This overlap might lead to slight inaccuracies in distinguishing between closely related TD types but is unlikely to significantly impact the overall validity of our conclusions.

\section{Conclusion}

In this study, we demonstrate advancements in identifying various TD types from issue trackers and project management systems. 
We found that fine-tuning transformer-based models on project specific TD data significantly enhances their performance. Our comparison between GPT and DistilRoBERTa models revealed that while large language models like GPT show promise, our fine-tuned DistilRoBERTa model often outperformed them. This highlights the importance of targeted fine-tuning over sheer model size in TD classification tasks.
Notably, our expert binary classifiers consistently outperformed multi-class approaches, providing more precise and reliable identification of specific TD types. This improvement is particularly evident in our ability to distinguish between closely related TD categories, a challenge in previous studies.
A key strength of our study is the extensive evaluation in both open-source and industrial settings. Our models demonstrated robust performance on out-of-distribution data, including a real-world dataset from a multinational software company. This industrial validation underscores the practical applicability of our approach in diverse and evolving software projects.
By releasing our curated dataset and presenting these improved methods, we aim to drive further advancements in TD classification research. Our work not only enhances the accuracy of TD detection but also provides a more nuanced understanding of different TD types, contributing to more effective software maintenance and quality assurance strategies in both academic and industrial contexts.

\bibliographystyle{IEEEtran}
\bibliography{sample-base}

\end{document}